\documentclass[preprintnumbers,amsmath,11pt,amssymb,floatfix,superscriptaddress,nofootinbib]{article}

\def\mytitle#1{\setcounter{equation}{0}
\setcounter{footnote}{0}
\begin{flushleft}\Large\textbf{#1}\end{flushleft}
\vspace{0.25cm}}
\def\myname#1{\leftline{{\large #1}}\vspace{-0.13cm}}
\def\myplace#1#2{\small\begin{flushleft}\textit{#1}\\
\texttt{#2}\end{flushleft}}

\def\myclassification#1{\small\noindent
Keywords :
       #1\vspace{0.5cm}}
\usepackage{graphicx}
\usepackage{amsmath,latexsym}
\begin{document}
\mytitle{Properties of Particle Trajectory Around a Weakly Magnetized Black Hole}

\myname{$Amritendu~ Haldar^{*}$\footnote{amritendu.h@gmail.com} and $Ritabrata~
Biswas^{\dag}$\footnote{biswas.ritabrata@gmail.com}}
\myplace{*Department of Physics, Sripat Singh College, Jiaganj, Murshidabad $-$ 742123, India.\\$\dag$ Department of Mathematics, The University of Burdwan, Golapbag Academic Complex, City : Burdwan  $-$ 713 104, District : Purba Barddhaman, State : WestBengal, India.} {}
 
\begin{abstract}
In this paper, we consider charged accelerating AdS black holes with nonlinear electromagnetic source. The metric chosen by us is of a regular black hole which shows regular nature at poles and a conical effect which corresponds to a cosmic string. In such a space time construction of the Lagrangian for a charged particle is done. Cyclic coordinates as well as the corresponding symmetry generators, i.e., the Killing vectors are found. Conservation laws corresponding to the symmetries are counted. Euler Lagrange equations are found. The orbit is mainly taken to be a circular one and effective potential is found. The minimum velocity obtained by a particle to escape from innermost stable circular orbit is found. The values of this escape velocity is plotted with respect to the radius of the event horizon of the central black hole for different parametric values. The nature of the escape velocity is studied when the central object is working with gravitational force and charge simultaneously. Effective potential and effective force are also plotted. The range of radius of event horizon for which the effective force turns to be positive is found out. A pathway of future studies of accretion disc around such black holes is made.          
\end{abstract}

\myclassification{Euler-Lagrange Equation; Cyclic Co-ordinate; Escape Velocity;  Center of Mass Energy and Effective Force.}

\section{Introduction}

Black Hole (BH hereafter) is the important prediction of General Relativity (GR hereafter). BHs are specified by the curved space-time geometry and bounded by the event horizons. Now a days, the study of dynamics of particles around the BHs in the astrophysical background is found in many literature. Both the particles and photons in the vicinity of the BHs are highly attracted by the strong gravitational pull and as a result the accreted particles cause the gain of mass of the BHs. Again it is also possible that the BHs throw the particles away with high relativistic velocity due to angular momentum barrier. Furthermore the highly energised charged particles may also be escaped from their stable orbits around the BHs due to the collisions with other particles and are moved under the influence of the Lorentz force in the electromagnetic fields.

If the gravitational and electromagnetic fields are strong, the motions of the charged particles become, in general, chaotic and undetermined and so the corresponding orbits become unstable except the inner most stable circular orbit. The weak magnetic field does not affect the geometry of the concerned central BH but it may affect the motion of charged particles \cite{Znajek 1976, Blandford 1977}. In the experimental point of view \cite{Frolov 2003, Borm 2013}, the magnetic field around the BHs occurs due to existence of plasma in the form of an accretion disc or a charged gas cloud \cite{Mckinney 2007, Dobbie 2008} surrounding the BHs.

Mechanically, when the particles of a system collide with each other, the Center of Mass Energy (CEM hereafter) occurs. Near the event horizon of a BH, the high  CEM is produced due to the collision of two particles. The mechanism of collision of two particles falling towards the Kerr BH has been proposed by Banados Silk and West (BSW mechanism hereafter) \cite{Bandos 2009}. Furthermore, the same authors have also shown that the CEM, in the equatorial plane, may be highest for a fastly rotating BH. The BSW mechanism has been investigated for different BHs \cite{Berti 2009, Patil 2010, Patil 2011, Wei1 2010,  Zaslavskii 2010, Harada 2011, Grib 2011, HussianI 2012, Sadeghi 2012, Bambi 2013, Tuesunov 2013}. A general review of the mechanism of collision is viewed in \cite{Harada 2014}. The CEM of the particles at the inner horizon of Kerr BH \cite{Lake 2010}, the CEM of the collision of the particles around Kerr- Newmann BH \cite{Wei2 2010}, near the horizon(s) of Kerr-Taub-NUT BH \cite{Liu 2011}, cylindrical BH \cite{Said 2011}, Plebanski-Demianski BH \cite{Sharif 2013}, Kerr-Newmann-Taub-NUT BH \cite{Zakria 2015}, charged dilaton BH \cite{Pradhan 2015} have been studied.

The study of the dynamics of particles around the BHs in the presence of magnetic field has a special significance in recent time. The authors of some literatures \cite{Forlov 2010, Forlov 2012, Zahrani 2013} have investigated the motion of a charged particle near weakly magnetized Schwarzschild BH. In some other literatures \cite{Nakamura 1993, Takahashi 2009, Kopacek 2010, Preti 2010, Forlov 2011, Igata 2011} the authors have analyzed the chaotic motion of a charged particles around Kerr BH near magnetic field. More over in another paper \cite{Pugliese 2011} the authors have studied the circular motion of charged particles around Reissner-Nordstrom BH. \cite{Majeed 2015} have investigated  the dynamics of charged particles around slowly rotating Kerr BH with magnetic field and also \cite{Jamil 2015} have discussed about the dynamics of particles around Schwarzschild BH in the presence of quintessence and magnetic field. Recently, \cite{Jawad 2016} have investigated the dynamics of particles around a regular BH surrounded by external magnetic field. Zhou, S. et al \cite{Zhou_added} have studied the geodesic structure of the Janis-Newman-Winicour space time. This space time contains a strong curvature naked singularity. Solving the geodesic equation and analysing the behavior of effective potential, they have shown all geodesic types of the test particles and photons. In 2009 Horava Lifshitz gravity was introduced. Timelike geodesic motion in Horava Lifshitz space time was studied in \cite{Chen_added}. Differently energised particles' motions are studied.

Motivated by the previous works, our objective in this paper is to apply the $Euler-Lagrange$ equations of motion to study the properties of particle trajectories around the charged accelerating AdS (Anti-de-Sitter) BH surrounded by weak magnetic field. 

This paper is organized as follows: in next section, we study a charged accelerating AdS BH metric with non-linear electromagnetic source and establish the $Euler-Lagrange$ equation of motion radially. In section 3, we analyze the dynamics of charged particles with and without magnetic field around a charged accelerating AdS BH. In section 4, we investigate the CME of two colliding charged particles in the aforesaid conditions and in section 5, we calculate the effective force and explain it with graphical representation. Finally in the last section, we present a brief conclusion of the work.            
\section{Charged Accelerating AdS (anti-de-Sitter) Black Holes With Non-linear Electromagnetic Source:}

A charged accelerating AdS BHs can be expressed by the metric as \cite{Griffith 2006, Hang 2016}:
\begin{equation}\label{ah2.equn1}
ds^2=\frac{1}{\Omega^2}\left[f(r)dt^2-\frac{dr^2}{f(r)}{-r^2}\left(\frac{d\theta^2}{g(\theta)}+{g(\theta)sin^2\theta\frac{d\phi^2}{K^2}}\right)\right] ,
\end{equation}
with the electromagnetic field tensor $F$ which is related to the gauge potential $B$ as: 
\begin{equation}\label{ah2.equn2} 
F=dB~~~~~,~~~~~~B=-\frac{q}{r}dt
\end{equation} 
and where
\begin{equation}\label{ah2.equn3}
f(r)=(1-A^2r^2)\left(1-\frac{2m}{r}+\frac{q^2}{r^2}\right)+\frac{r^2}{l^2}~~,~~ g(\theta)=1+2mAcos\theta +q^2A^2cos^2\theta~~,~~\Omega(a~ conformal ~factor)=1+Arcos\theta~~ .
\end{equation} 
The factor $\Omega$ determines the conformal infinity or boundary of the AdS space-time. In the above expression (\ref{ah2.equn3}), $m$, $q$, $l=\sqrt{-\frac{\Lambda}{3}}$( where $\Lambda$ is cosmological constant) and $A$ represent the mass, electric charge of BHs, AdS radius and magnitude of acceleration of the BHs respectively. By looking at the angular part of the metric and the behavior of $g(\theta)$ at the both poles $\theta_+=0 $ and $\theta_-=\pi$, one can discover the presence of ``cosmic string". The regularity of the metric at a pole requires,
\begin{equation}\label{ah2.equn4}
K\pm=g(\theta\pm)=1\pm2Am+q^2A^2~~.
\end{equation}
We assume that the BH is regular on the North Pole $(\theta=0)$ with $K=1+2Am+q^2A^2$ and on the South Pole $(\theta=\pi)$ with $K=1-2Am+q^2A^2$ but there is a ``conical defect" which is $\delta=2\pi \left(1-\frac{g_-}{K_+}\right)=\frac{8\pi mA}{1-2Am+q^2A^2}$, which corresponds to a ``cosmic string" with tension $\mu =\frac{\delta}{8\pi}=\frac{\pi mA}{1-2Am+q^2A^2}$.
\section{Dynamics of Charged Particles:}
\subsection{In the Absence of Magnetic Field:}
The Lagrangian of a charged particle of mass $m$ and charge $q$ in the absence of magnetic field is expressed as \cite{Zahrani 2013, Jamil 2015}:
\begin{equation}\label{ah2.equn5}
{\cal L}=\frac{1}{2} g_{\mu\nu}\frac{dx^\mu}{d\tau}\frac{dx^\nu}{d\tau}.
\end{equation}
By employing the Lagrangian dynamics, we now examine the motion of that particle in the background of the BH (\ref{ah2.equn1}). The Lagrangian of that particle is as follows:
\begin{equation}\label{ah2.equn6}
{\cal L}(r)=\frac{1}{2\Omega^2}\left[f(r)\dot{t}^2-\frac{\dot{r}^2}{f(r)}{-r^2}\left(\frac{\dot{\theta}^2}{g(\theta)}+{g(\theta)sin^2\theta\frac{\dot{\phi}^2}{K^2}}\right)\right].
\end{equation}
The over dot indicates the derivative with respect to proper time. It is evident from the Lagrangian (\ref{ah2.equn6}) that $t$, $\theta$ and $\phi$ are cyclic coordinates and hence it leads to corresponding symmetry generators which are also known as Killing vectors. The metric (\ref{ah2.equn1}) remains invariant under the Killing vector fields $X$ and we obtain:
\begin{equation}\label{ah2.equn7}
X_0=\frac{\partial \cal L}{\partial \dot{t}}=\frac{f(r) \dot{t}}{\Omega^2},
\end{equation}
\begin{equation}\label{ah2.equn8}
X_1=\frac{\partial {\cal L}}{\partial \dot{\phi}}=-\frac{g(\theta)r^2 sin^2\theta \dot{t}}{\Omega^2}\frac{\dot{\phi}}{K^2},
\end{equation}
\begin{equation}\label{ah2.equn9}
X_2=cos\phi\frac{\partial {\cal L}}{\partial \dot{\theta}}-cot\theta sin\phi\frac{\partial {\cal L}}{\partial \dot{\phi}}=-\frac{r^2}{\Omega^2}\left(cos\phi \frac{\dot{\theta}}{g(\theta)}-g(\theta)sin\theta cos\theta sin\phi\frac{\dot{\phi}}{K^2}\right)
\end{equation}
and
\begin{equation}\label{ah2.equn10}
X_3=sin\phi\frac{\partial {\cal L}}{\partial \dot{\theta}}+cot\theta cos\phi\frac {\partial {\cal L}}{\partial \dot{\phi}}=-\frac{r^2}{\Omega^2}\left(sin\phi \frac{\dot{\theta}}{g(\theta)}+g(\theta)sin\theta cos\theta cos\phi\frac{\dot{\phi}}{K^2}\right)
\end{equation}
There are three commuting integral of motion corresponding $\it{eq^n}$ (\ref{ah2.equn1}) in which two of them are generated by the Killing vectors
\begin{equation}\label{ah2.equn11}
\xi_{(t)}=\xi_{(t)}^\mu=\partial_t~,~~~~~\xi_{(\phi)}=\xi_{(\phi)}^\mu=\partial_\phi~,
\end{equation} 
where $\xi_{(t)}^\mu=(1, 0 ,0 ,0)$ and $\xi_{(\phi)}^\mu=(0, 0, 0, 1),$ reflecting invariance under the time translation and rotation around the symmetry axis. The corresponding  conserved quantities, viz., specific energy $E$ and specific azimuthal angular momentum $L_z$ are given by,
\begin{equation}\label{ah2.equn12}
E\equiv-\frac{p_\mu \xi_{(t)}^\mu }{m}=\frac{\partial \cal L}{\partial \dot{t}}=\dot{t}\frac{f(r) }{\Omega^2}~,~~L_z\equiv \frac{p_\mu \xi_{(\phi)}^\mu }{m}=\frac{\partial {\cal L}}{\partial \dot{\phi}}=\dot{\phi}\frac{g(\theta)r^2 sin^2\theta }{\Omega^2 {K^2}}, 
\end{equation}
where $p_\mu=mu_\mu$ is the linear momentum of the particle in the absence of magnetic field.
And the other conservation laws corresponding to the symmetries are depicted as:\\
angular momenta,
\begin{equation}\label{ah2.equn13}
L_1=\frac{r^2}{\Omega^2}\left(cos\phi \frac{\dot{\theta}}{g(\theta)}-\frac{1}{2} g(\theta)sin2\theta sin\phi\frac{\dot{\phi}}{K^2}\right)~~~and~~~L_2=\frac{r^2}{\Omega^2}\left(sin\phi \frac{\dot{\theta}}{g(\theta)}+\frac{1}{2}g(\theta)sin2\theta  cos\phi\frac{\dot{\phi}}{K^2}\right).
\end{equation} 
It is obvious from $\it{eq^n}s$ (\ref{ah2.equn9}) and (\ref{ah2.equn10}) that if the particle which moves in the equatorial plane, $X_2$ and $X_3$ become irrelevant.

Here we apply the $\it{Euler-Lagrange} $ equation of motion for $r$ only and we obtain:
\begin{equation}\label{ah2.equn14}
\ddot{r}=\frac{1}{2}\left[2rf(r)\left(\frac{\dot{\theta}^2}{g(\theta)}+g(\theta) sin^2\theta\frac{\dot{\phi}^2}{K^2}\right)-f(r)f^\prime(r)\dot{t}^2-\dot{r}^2\frac{f^\prime(r)}{f(r)}\right],
\end{equation} 
where $f(r)$, $\dot{t}$ and $\dot{\phi}$ are obtained from $\it{eq^n}s$ (\ref{ah2.equn3}) and (\ref{ah2.equn12}).

The constant of motion for that particle is given by  $\it{eq^n}$ (\ref{ah2.equn12}). The particle is associated with the total specific angular momentum as \cite{Frolov 2003,  Jawad 2016}:
\begin{equation}\label{ah2.equn15}
L^2\equiv r^4\dot{\theta}^2+\frac{L_z^2\Omega^2 K^4}{g(\theta)^2 sin^2\theta}= r^2v_{\perp}+\frac{L_z^2\Omega^2 K^4}{g(\theta)^2 sin^2\theta} ,
\end{equation}
Applying the normalization condition for four-velocity, we calculate $\dot{r}^2$ from $\it{eq^n}$ (\ref{ah2.equn6}) as:
\begin{equation}\label{ah2.equn16}
\dot{r}^2=f(r)^2 \dot{t}^2-f(r)\left[\Omega^2+\frac{r^2\dot{\theta}^2}{g(\theta)}+{g(\theta)r^2 sin^2\theta\frac{\dot{\phi}^2}{K^2}})\right].
\end{equation}
Since we assume the system is spherically symmetric, all $\theta=$  constant planes will be equivalent to the equatorial plane for which $\theta=\frac{\pi}{2} $ and hence from $\it{eq^n}$ (\ref{ah2.equn16}) we have
\begin{equation}\label{ah2.equn17}
\dot{r}^2=E^2 -f(r)\left[1+\frac{L_z^2 K^2}{r^2}\right].
\end{equation}
Again if we consider the orbit of the particle through which it is moving is circular, then the $\it{eq^n}$ (\ref{ah2.equn17}) reduces to 
\begin{equation}\label{ah2.equn18}
E^2 =f(r)\left[1+\frac{L_z^2 K^2}{r^2}\right]
\end{equation}
and it is equivalent to the effective potential $U_{eff}(r)$ of that particle. Hence $\it{eq^n}$ (\ref{ah2.equn18}) shows that the total energy and / or the effective potential will vanish at the horizon(s).
 
The critical azimuthal angular momentum of a particle that follows a particular orbit where the effective potential is extremum (i.e., maximum or minimum) and it is expressed as:
\begin{equation}\label{ah2.equn19}
L_z^2=\frac{g(\theta)}{K^2}\left(\frac{mr-q^2+A^2mr^3-A^2r^4+\frac{r^4}{l^2}-2\frac{A^2r^6}{l^2}}{1+A^2mr-A^2q^2-3\frac{m}{r}+2\frac{q^2}{r^2}+\frac{A^2r^2}{l^2}}\right).
\end{equation}
Hence the energy of the particle would be 
\begin{equation}\label{ah2.equn20}
E^2=\frac{f(r)^2}{1+A^2mr-A^2q^2-3\frac{m}{r}+2\frac{q^2}{r^2}+\frac{A^2r^2}{l^2}}.
\end{equation}
After collision, the energy of the particle takes the form as:
\begin{equation}\label{ah2.equn21}
E_c^2=f(r)\left[1+\frac{\left(L_zK+rv_\perp\right)^2}{r^2}\right].
\end{equation}
So comparing between the $\it{eq^ns}$ (\ref{ah2.equn18}) and (\ref{ah2.equn21}) we infer that the energy after collision is grater than that of the total energy before collision as the extra term $rv_\perp$ is present in $\it{eq^n}$ (\ref{ah2.equn21}) and which is obvious due to the collision. In this expression $v_\perp$ is the minimum velocity of the particle required to escape from innermost stable circular orbit (ISCO hereafter) and is given as:
\begin{equation}\label{ah2.equn22}
v_\perp=\sqrt{{\frac{E_c^2-f(r)}{f(r)}}}-{\frac{L_zK}{r\sqrt{g(\theta)}}}.
\end{equation}  
\subsection{In the Presence of Magnetic Field:}
We now consider the case of weakly magnetized BH and investigate the motion of a charged particle having charge $q$ in presence of magnetic field in the BH exterior. The general Killing vector equation is \cite{Wald 1974}
\begin{equation}\label{ah2.equn23}
\Box\xi^\mu=0,
\end{equation}
where $\xi^\mu$ is a killing vector  and this equation coincides with the Maxwell equation for 4-potential $A^\mu$ in the Lorentz gauge  $A_{~;\mu}^\mu =0$. The special choice \cite{Aliev 1978}
\begin{equation}\label{ah2.equn24}
A^\mu=\frac{\cal B}{2}\xi_{(\phi)}^\mu~,
\end{equation}
corresponds to the test magnetic field, where $\cal B$ is magnetic field strength. The 4-potential discussed in the $eq^n (\ref{ah2.equn24})$ is invariant under the symmetries corresponding to the Killing vectors as discussed above, i.e.,
\begin{equation}\label{ah2.equn25}
L_\xi A_\mu=A_{\mu, \nu}\xi^\nu-A_\nu\xi_{,\mu}^\nu=0.
\end{equation}
A magnetic field vector with respect to an observer whose 4-velocity is $u^\mu$, defined as:\\
\begin{equation}\label{ah2.equn26}
{\cal B}=-\frac{1}{2} e^{\mu \nu \lambda \sigma} {F_{\lambda \sigma}} u_\nu~,
\end{equation}
where $e^{\mu \nu \lambda \sigma}=\frac{\epsilon^{\mu \nu \lambda \sigma}}{\sqrt{-g}}~, \epsilon_{0123}=1$ and $g=det(g_{\mu \nu}).$ $\epsilon^{\mu \nu \lambda \sigma}$ is the Levi symbol and $F_{\lambda \sigma}$ is the Maxwell tensor, which is given by 
\begin{equation}\label{ah2.equn27}
F_{\lambda \sigma}=A_{\sigma,\lambda}-A_{\lambda,\sigma}~~.
\end{equation}\\
The Lagrangian of a particle of mass $m$ and charge $q$ in the presence of magnetic field is expressed by \cite{Zahrani 2013, Jamil 2015}:
\begin{equation}\label{ah2.equn28}
{\cal L}=\frac{1}{2} g_{\mu\nu}\frac{dx^\mu}{d\tau}\frac{dx^\nu}{d\tau}+\frac{q}{m} A_\mu \frac{dx^\mu}{d\tau},
\end{equation}
where $A_\mu$ is the 4-vector potential for the electromagnetic field.
 
By employing the Lagrangian dynamics, we now examine the motion of a charged particle in the background of the BH (\ref{ah2.equn1}). Here the Lagrangian of that particle is as follows:
\begin{equation}\label{ah2.equn29}
{\cal L}(r)=\frac{1}{2\Omega^2}\left[f(r)\dot{t}^2-\frac{\dot{r}^2}{f(r)}{-r^2}\left(\frac{\dot{\theta}^2}{g(\theta)}+{g(\theta)sin^2\theta\frac{\dot{\phi}^2}{K^2}}\right)\right]+\frac{q\cal B}{2m}\frac{g(\theta)r^2sin^2\theta}{\Omega^2}\frac{\dot{\phi}}{K^2}.
\end{equation}
The generalized 4-momentum of the particle is denoted as, 
\begin{equation}\label{ah2.equn30}
P_\mu= mu_\mu+qA_\mu.
\end{equation} 
Then the new conservation laws corresponding to the symmetries are defined below
\begin{equation}\label{ah2.equn31}
E\equiv-\frac{P_\mu \xi_{(t)}^\mu }{m}=\frac{\partial \cal L}{\partial \dot{t}}=\dot{t}\frac{f(r) }{\Omega^2}~,~~L_z\equiv \frac{P_\mu \xi_{(\phi)}^\mu }{m}=\frac{\partial \cal L}{\partial \dot{\phi}}=\left(\dot{\phi}+\frac{q\cal B}{2m}\right)\frac{g(\theta)r^2 sin^2\theta }{\Omega^2 {K^2}}, 
\end{equation}
Here we denote $B\equiv\frac{q\cal B}{2m}$. Hence the new constants of motion will be given by:
\begin{equation}\label{ah2.equn32}
\dot{t}=\frac{E\Omega^2}{f(r)}~,~~ \dot{\phi}=\frac{L_z\Omega^2 K^2}{g(\theta) r^2 sin^2\theta}{-B}.
\end{equation}
The equation of motion of the charged particle in this case is obtained by applying by the $\it{Euler-Lagrange} $ equation, which will be of the form as:
\begin{equation}\label{ah2.equn33}
\ddot{r}=\frac{1}{2}\left[2rf(r)\left(\frac{\dot{\theta}^2}{g(\theta)}+g(\theta) sin^2\theta\frac{\dot{\phi}^2}{K^2}+2Bg(\theta) sin^2\theta\frac{\dot{\phi}}{K^2}\right)-f(r)f^\prime(r)\dot{t}^2-\dot{r}^2\frac{f^\prime(r)}{f(r)}\right],
\end{equation}
where $f(r)$, $\dot{t}$ and $\dot{\phi}$ are obtained from $\it {eq^n}$ (\ref{ah2.equn32}). 
Applying the normalisation condition for four-velocity in $\it{eq^n}$ (\ref{ah2.equn29}) and using $\it{eq^n}$ (\ref{ah2.equn32}), we have the energy and the corresponding effective potential as:
\begin{equation}\label{ah2.equn34}
E^2=\dot{r}^2+f(r)\frac{r^2\dot{\theta}^2}{g(\theta)}+f(r)\left[\Omega^2+g(\theta)r^2sin^2\theta\frac{1}{K^2}\left(\frac{L_z\Omega^2 K^2}{g(\theta) r^2 sin^2\theta}{-B}\right)^2\right]
\end{equation}
and~~~
\begin{equation}\label{ah2.equn35}
U_{eff}(r)=f(r)\left[\Omega^2+g(\theta)r^2sin^2\theta\frac{1}{K^2}\left(\frac{L_z\Omega^2 K^2}{g(\theta) r^2 sin^2\theta}{-B}\right)^2\right].
\end{equation}
For the condition as applied in $\it {eq^n}$ (\ref{ah2.equn16}), we have from $\it {eq^ns}$ (\ref{ah2.equn34}) and (\ref{ah2.equn35}) that 
\begin{equation}\label{ah2.equn36}
E^2=\dot{r}^2+f(r)\left[1+\frac{r^2}{K^2}\left(\frac{L_z K^2}{ r^2 }{-B}\right)^2\right]
\end{equation}
and~~~
\begin{equation}\label{ah2.equn37}
U_{eff}(r)=f(r)\left[1+\frac{r^2}{K^2}\left(\frac{L_z K^2}{ r^2}{-B}\right)^2\right].
\end{equation}
\subsection{Dimensionless form of the Equations:}
In order to integrate the dynamical equations, we need to make these equations dimensionless. We use the following transformation relations \cite{Jamil 2015, Hsu 2004, HussainS 2014} as:

\begin{equation}\label{ah2.equn38}
2m=r_d,~~ \tau=\sigma r_d,~~ r=\rho r_d,~~ L_z=\ell r_d,~~ q=\alpha r_d,~~ B=b r_d,~~ l=s r_d~~ and~~ a=A r_d.
\end{equation}
Using these relations (\ref{ah2.equn38}), the $\it{eq^ns}$ (\ref{ah2.equn33}), (\ref{ah2.equn34}) and (\ref{ah2.equn35}) acquire the form as:
$$\ddot{\rho}=\frac{1}{2}\left[2\rho f(\rho) \left(\frac{ \dot{\theta}^2}{g(\theta)}+g(\theta)sin^2\theta\frac{\dot{\phi}^2}{K^2}+2bg(\theta) sin^2\theta\frac{\dot{\phi}}{K^2}\right)-f(\rho)f^\prime(\rho)\dot{t}^2-\frac{\dot{\rho}^2 f^\prime(\rho)}{f(\rho)}\right],$$
$$E^2=\dot{\rho}^2+f(\rho)\frac{\rho^2 \dot{\theta}^2}{g(\theta)}+f(\rho)\left[\Omega^2+g(\theta)\rho^2sin^2\theta\frac{1}{K^2}\left(\frac{\ell\Omega^2 K^2}{g(\theta) \rho^2 sin^2\theta}{-b}\right)^2\right]$$
and
\begin{equation}\label{ah2.equn39}
U_{eff}(\rho)=f(\rho)\left[\Omega^2+g(\theta)\rho^2sin^2\theta\frac{1}{K^2}\left(\frac{\ell\Omega^2 K^2}{g(\theta) \rho^2 sin^2\theta}{-b}\right)^2\right],
\end{equation}
where~~$$f(\rho)=(1-a^2\rho^2)\left(1-\frac{1}{\rho}+\frac{\alpha^2}{\rho^2}+\frac{\rho^2}{s^2} \right)$$ ~and~
\begin{equation}\label{ah2.equn40}
\Omega=1+a\rho cos\theta.
 \end{equation}
After collision, for $\theta=\frac{\pi}{2}$ and constant $\rho$ the energy given in $\it{eq^n}$ (\ref{ah2.equn39}) reduces to 
\begin{equation}\label{ah2.equn41}
E_c^2=f(\rho)\left[1+\frac{\rho^2}{K^2}\left(\frac{\ell K^2+\rho v_\perp}{\rho^2}-{b}\right)^2\right].
\end{equation}
Hence the escape velocity of the charged particle is expressed as:
\begin{equation}\label{ah2.equn42}
v_\perp=K\sqrt{{\frac{E_c^2-f(\rho)}{f(\rho)}}}-{\frac{\ell K^2}{\rho}}+\rho b.
\end{equation}
\begin{figure}[h!]
\begin{center}
~~~~~~Fig.-1a ~~~~~~~~~~~~~~~~~~~~Fig.-1b~~~~~~~~~~~~~~~~~~~\\~~\\
\includegraphics[scale=.8]{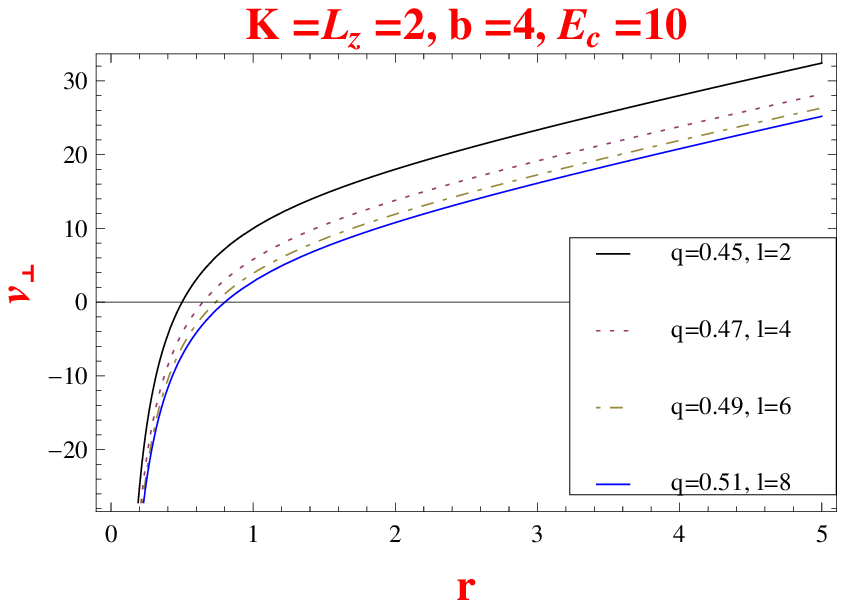}~~
\includegraphics[scale=.8]{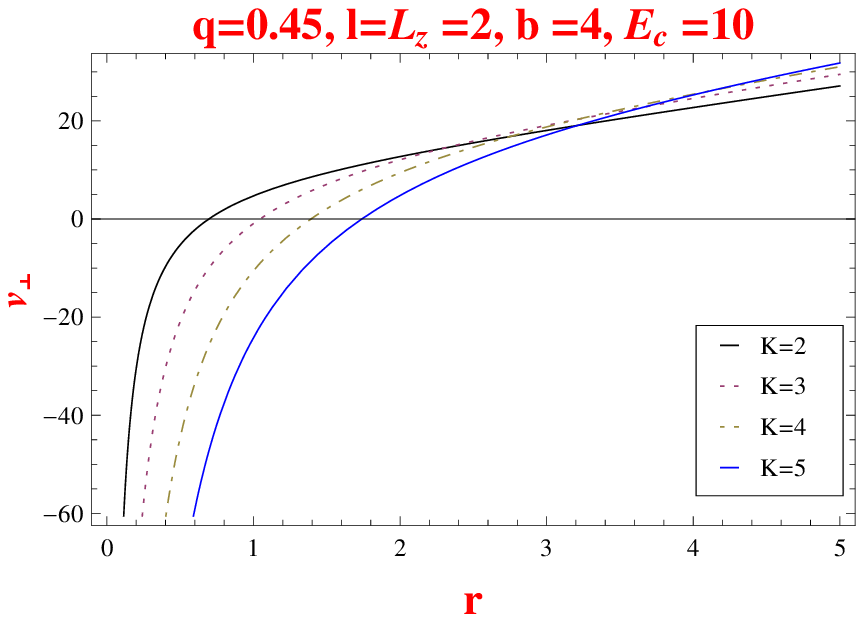}\\
Fig-$ 1a $ represents the variation of escape velocity $ v_\perp $  with respect to $ r $ keeping $K$, $ L_z $ and $b$ are fixed, with varying $q$ and $l$, \\
Fig-$ 1b $ represents the variation of escape velocity $ v_\perp $  with respect to $ r $ keeping $ q $, $ l $, $ L_z $ and   $ b $ are fixed with varying $K$,\\~~\\
~~~~~~Fig.-1c ~~~~~~~~~~~~~~~~~~~~Fig.-1d~~~~~~~~~~~~~~~~~~~\\~~\\
\includegraphics[scale=.8]{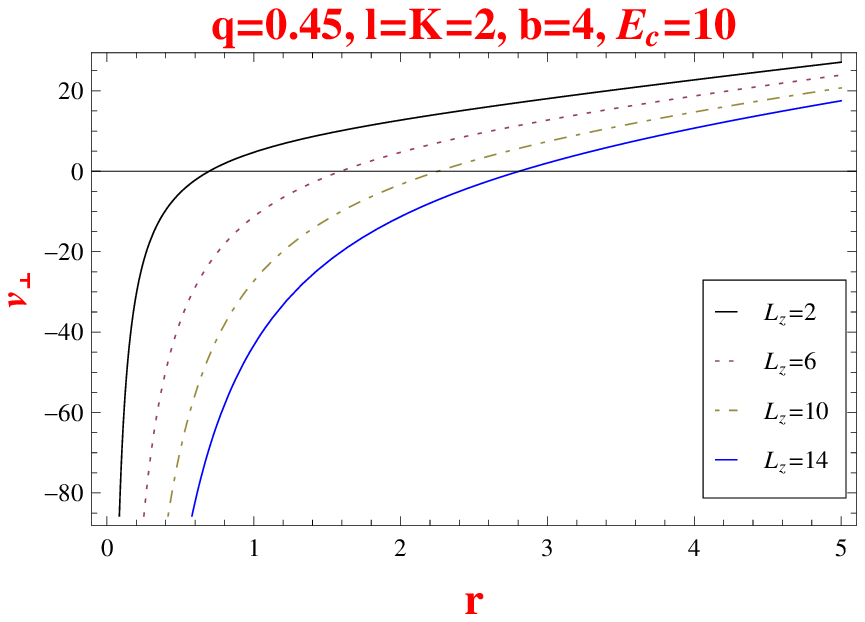}~~
\includegraphics[scale=.8]{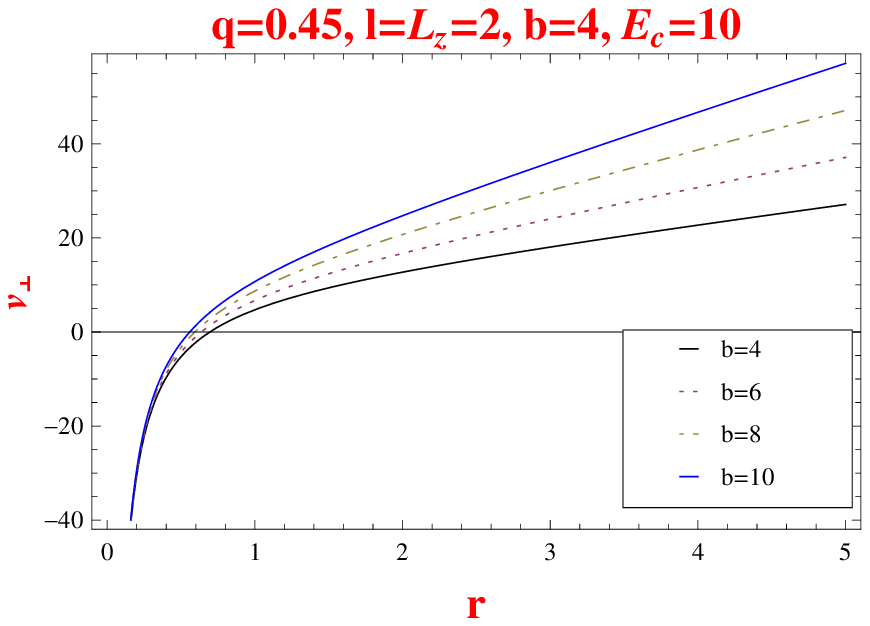}\\
Fig-$ 1c $ represents the variation of escape velocity $ v_\perp $  with respect to $ r $ keeping $ q $, $ l $, $ K $  and $b$  are fixed with varying $ L_z $,\\
Fig-$ 1d $ represents the variation of escape velocity $ v_\perp $  with respect to $ r $ keeping $ q $, $ l $, $ L_z $ and   $ K $ are fixed with varying $b$, \\
\end{center} 
\end{figure}

\begin{figure}[h!]
\begin{center}
~~~~~~~~Fig.-1e~~~~~~~~~\\~~\\
\includegraphics[scale=.8]{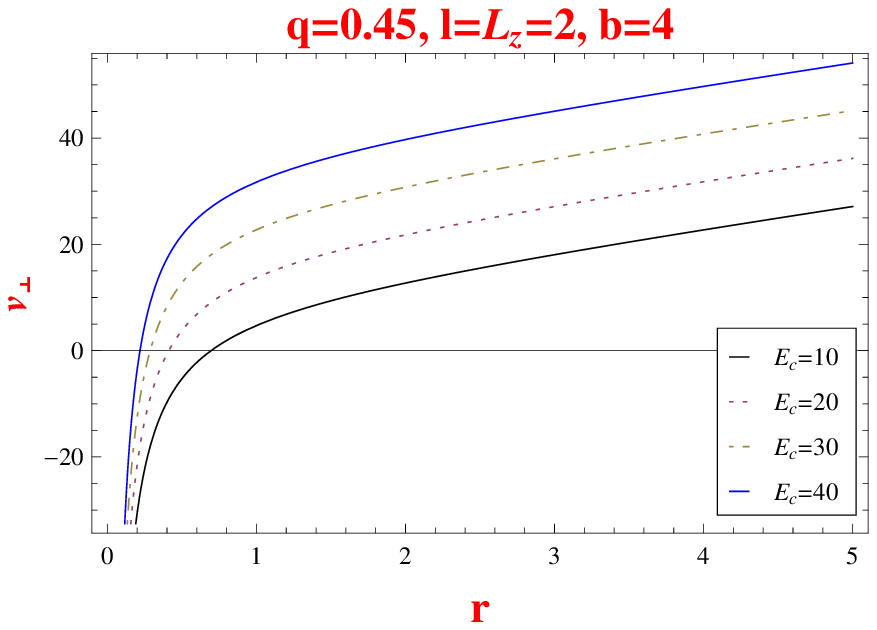}\\
Fig-$ 1e $ represents the variation of escape velocity $ v_\perp $  with respect to $ r $ keeping $K$, $ L_z $ $q$, $l$ and $b$ are fixed, with varying $E_c$.\\
\end{center} 
\end{figure}
We have plotted the variation of escape velocity $ v_\perp $ of the particle with respect to the radius $r$ of the BH  for different values of $q, l, K, b, L_z$ and $E_c$ in Fig- $(1a - 1e)$.

In Fig-1a, we vary $q$ and $l$ simultaneously keeping $K, b, L_z$, and $E_c$ fixed and we notice that with increment of $q$ and $l$, for same $r$, the escape velocity $ v_\perp $ decreases. If $r$ is small enough, $v_{\perp}$ has a negative value which depicts no physical particle can escape from so near points of a compact object considered here. But once distance from the center of the gravitating object is made higher, we can have an escape velocity which firstly will increase steeply with the increment of $r$ and latter the rate of increment will be reduced. However, the value is ever increasing with $r$.  If we increase $q$ and $l$ keeping all the other parameters constant, the curves stays of the same nature only except the fact that they are amplified. This means, for a higher charge, escape velocity is lower. So is for the higher space time curvature, i.e., higher the curvature lower is the escape velocity. This phenomenon is quite obvious as whenever the central engine is attracting gravitationally and electrically it is more hard for the particle to escape from such an object. Besides, if the curvature is high, higher velocity is required.

In Fig-1b, we change the values of $K$ by fixing the other parameters and we find that with increase of $K$, the particle escapes from higher values of $r$ but for same $r$, the escape velocity $ v_\perp $ increases. As we vary $K$ only again for low $r$ we see unphysical escape velocity. With $r$ escape velocity increases. For low $r$, low $K$ a high $v_{\perp}$ will be required. For high $r$, low $K$, the $v_{\perp}$ required is low. For every two $K_{low}$ and $K_{high}$ curves there is a point $r_{crit}$ where the system needs same escape velocity. This is due to the quadratic nature of $K$ in the expression of $v_{\perp}$ in $\it{eq^n}$ (\ref{ah2.equn42}).

We vary $L_z$ keeping the parameters $q, l, K, b,$, and $E_c$ are unchanged in Fig-1c and we observed that its nature is similar as Fig-1b. Here we noticed that with increase of $L_z$, the particle escapes from higher values of $r$ but for same $r$, the escape velocity $ v_\perp $ decreases.

Fig-1d and Fig-1e have the similar nature. One is plotted by varying the strength of magnetic field $b$ and other with escape energy $E_c$. We observe from the figures that for higher strength of magnetic field as well as escape energy the escape velocity of the particle is also higher. Hence we conclude that the magnetic field which escape the particle from the vicinity of the BH plays an important role in transfer mechanism of energy.
\begin{figure}[h!]
\begin{center}
~~~~~~Fig.-2a ~~~~~~~~~~~~~~~~~~~~Fig.-2b~~~~~~~~~~~~~~~~~~~\\~~\\
\includegraphics[scale=.8]{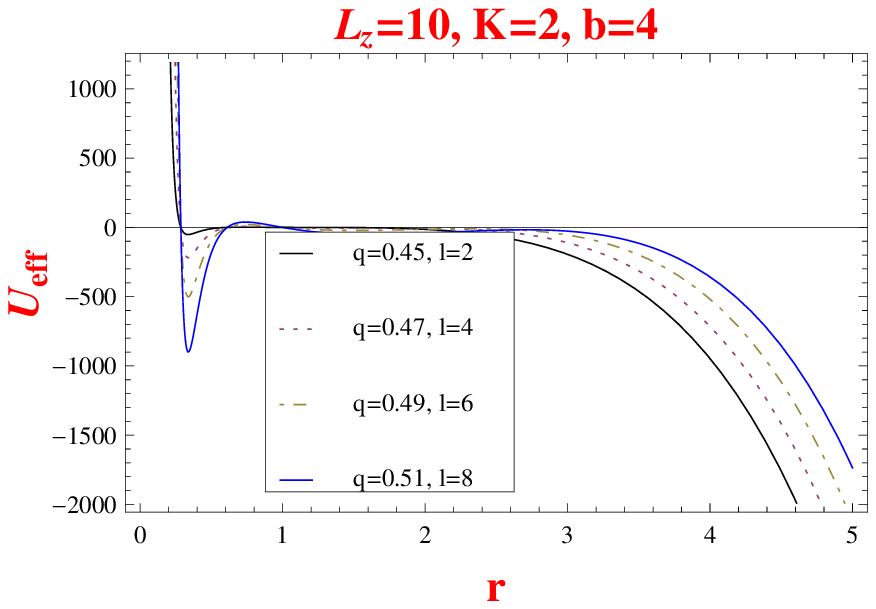}
\vspace{.6cm}
\includegraphics[scale=.8]{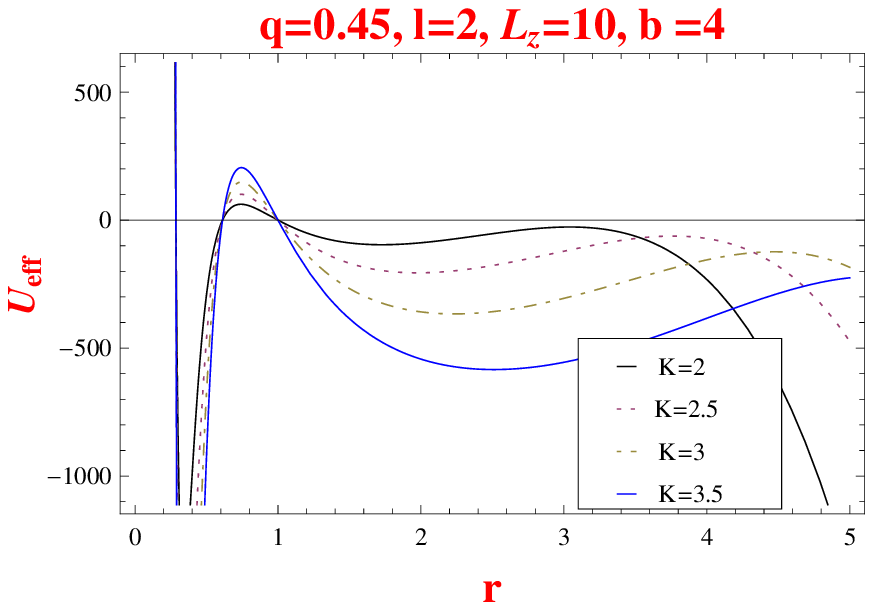}\\
Fig-$ 2a $ represents the variation of effective potential $ U_{eff} $  with respect to $ r $ keeping $K$, $ L_z $ and $b$ are fixed, with varying $q$ and $l$. \\
Fig-$ 2b $ represents the variation of effective potential $ U_{eff} $  with respect to $ r $ keeping $ q $, $ l $, $ L_z $ and $ b $ are fixed with varying $K$. \\~~\\
~~~~~~Fig.-2c ~~~~~~~~~~~~~~~~~~~~Fig.-2d~~~~~~~~~~~~~~~~~~~\\~~\\
\includegraphics[scale=.8]{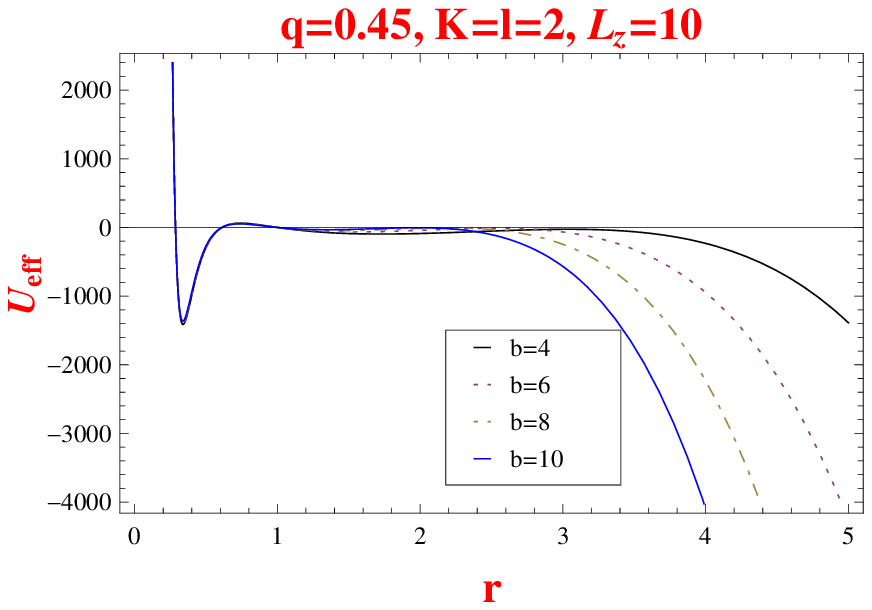}
\vspace{.6cm}
\includegraphics[scale=.8]{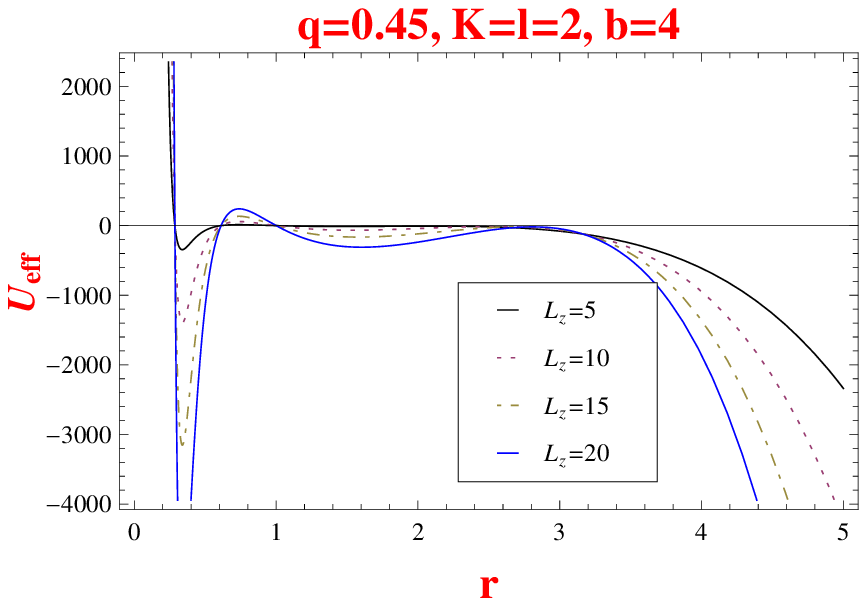}\\
Fig-$ 2c $ represents the variation of effective potential $ U_{eff} $  with respect to $ r $ keeping $ q $, $ l $, $ L_z $ and $ K $ are fixed with varying $b$. \\
Fig-$ 2d $ represents the variation of effective potential $ U_{eff} $  with respect to $ r $ keeping $ q $, $ l $, $ K $  and $b$  are fixed with varying $ L_z $.
\end{center}
\end{figure}

We have drawn the curves of effective potential $U_{eff}$ of the particle corresponding to circular orbits with respect to radius $r$ of BH for different values of $q, l, K, b, $ and $L_z$ in Fig- $(2a - 2d)$. We observe a similar feature that there is a minimum value of $U_{eff}$ at $r$ lying between 1 and 2. Initially $U_{eff}$ decreases very sharply (almost straight down) with increases of $r$ and gets negative value when $r$ exceeds the value about 1 and it reaches to a minimum value. Further increases of $r$, $U_{eff}$ increases to nearly zero and then rapidly decreases.

In Fig-$2a$, it is found that $U_{eff}$ gets nearly zero value at about $r=3$ for all the variation of $q$ and $l$. But Fig-$(2b-2d)$ show that $U_{eff}$ reaches nearly zero value for different values of $r$. Due to increase of corresponding parameters, the values of $r$ where $U_{eff}$ reaches nearly zero value also increase.     

\section{Center of Mass Energy of the Colliding Charged Particles:}
The CME of the colliding particles is expressed as \cite{Jawad 2016}:
\begin{equation}\label{ah2.equn43}
E_{CME}=\sqrt{2}m_o\left(1-g_{\mu\nu}u^\mu u^\nu \right)^{\frac{1}{2}},
\end{equation}
where $m_o$ is the mass and $u^\mu$ is the 4-velocity of each particles respectively.
\subsection{In the Absence of Magnetic Field:}
Applying $\it{eq^n}$ (\ref{ah2.equn12}) in $\it{eq^n}$ (\ref{ah2.equn43}), we obtain 

\begin{equation}\label{ah2.equn44}
E_{CME}=\sqrt{2}m_o\frac{L_z}{r}\left(1-K^2 \right)^{\frac{1}{2}}\simeq\sqrt{2}m_o\frac{L_z}{r}\left(1-\frac{K^2}{2} \right),
\end{equation}

\subsection{In the Presence of Magnetic Field:}

Applying $\it{eq^n}$ (\ref{ah2.equn32}) in $\it{eq^n}$ (\ref{ah2.equn43}), we obtain 
\begin{equation}\label{ah2.equn45}
E_{CME}=\sqrt{2}m_o\frac{L_z}{r}\left(1-K^2-\frac{r^4B^2}{L_z^2}+2\frac{r^4B}{L_z} \right)^{\frac{1}{2}}\simeq \sqrt{2}m_o\frac{L_z}{r}\left(1-\frac{K^2}{2}-\frac{r^4B^2}{2L_z^2}+\frac{r^4B}{L_z}  \right),
\end{equation}
It is evident from $\it{eq^ns}$ (\ref{ah2.equn44}) and (\ref{ah2.equn45}) that the CEM of the charged particles with and without magnetic field do not change at horizon(s).

\section{Effective Force :}
The effective force acting on the charged particles in the flat background normally measured by the Lorentz force but in the curved background it may be determined as follows 
$$F_{eff}=-\frac{1}{2}\frac{dU_{eff}}{dr} =a^2r\left(1-\frac{1}{r}+\frac{\alpha^2}{r^2}+\frac{r^2}{s^2}\right) \left[1+\frac{r^2}{K^2}\left(\frac{\ell K^2}{r^2}-{b}\right)^2\right]$$ 
\begin{equation}\label{ah2.equn46}
 - \left(1-a^2r^2\right)\left(\frac{1}{2r^2}-\frac{\alpha^2}{r^3}+\frac{r}{s^2}\right)\left[1+\frac{r^2}{K^2}\left(\frac{\ell K^2}{r^2}-{b}\right)^2\right]+ 4\frac{\ell}{r^2}\left(1-a^2r^2\right)\left(1-\frac{1}{r}+\frac{\alpha^2}{r^2}+\frac{r^2}{s^2}\right)\left(\frac{\ell K^2}{r^2}-{b}\right). 
 \end{equation} 
\begin{figure}[h!]
\begin{center}
~~~~~~Fig.-3a ~~~~~~~~~~~~~~~~~~~~Fig.-3b~~~~~~~~~~~~~~~~~~~\\~~\\
\includegraphics[scale=.8]{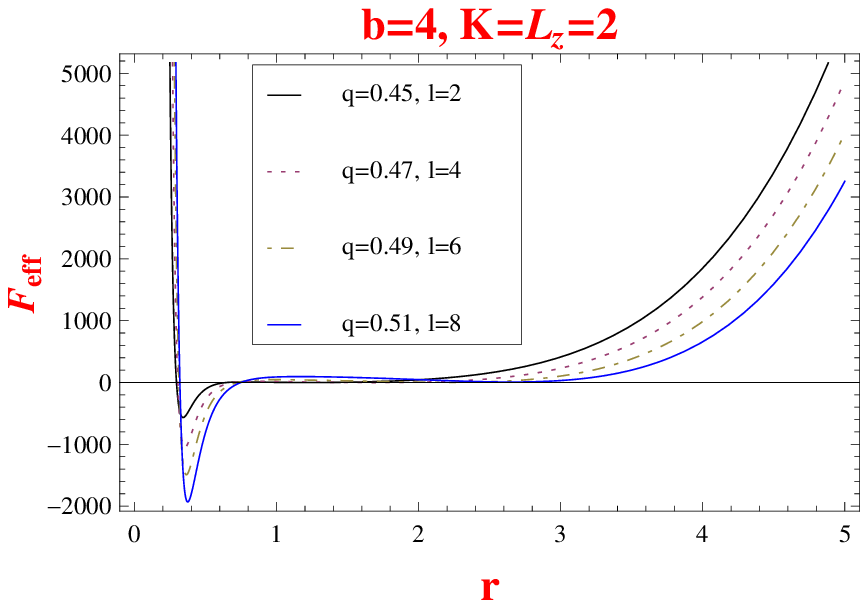}
\vspace{.6cm}
\includegraphics[scale=.8]{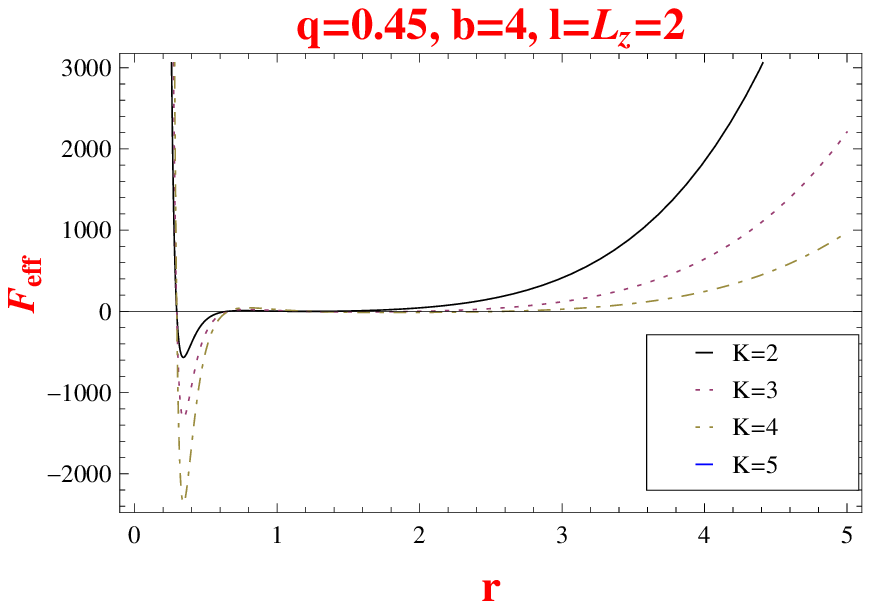}\\
Fig-$ 3a $ represents the variation of effective force $ f_{eff} $  with respect to $ r $ keeping $K$, $ L_z $ and $b$ are fixed, with varying $q$ and $l$. \\
Fig-$ 3b $ represents the variation of effective force $ f_{eff} $  with respect to $ r $ keeping $ q $, $ l $, $ L_z $ and   $ b $ are fixed with varying $K$. \\~~\\
~~~~~~Fig.-3c ~~~~~~~~~~~~~~~~~~~~Fig.-3d~~~~~~~~~~~~~~~~~~~\\~~\\
\includegraphics[scale=.8]{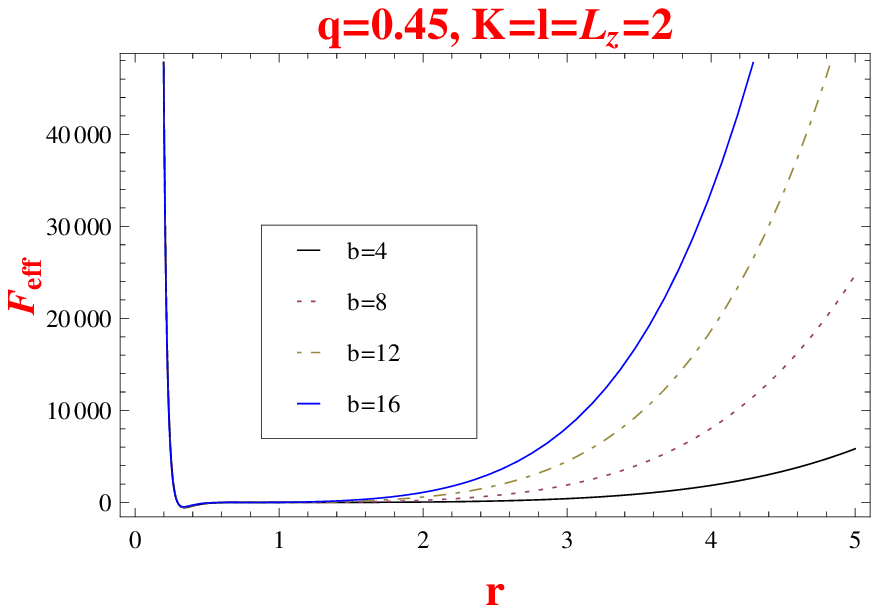}
\vspace{.6cm}
\includegraphics[scale=.8]{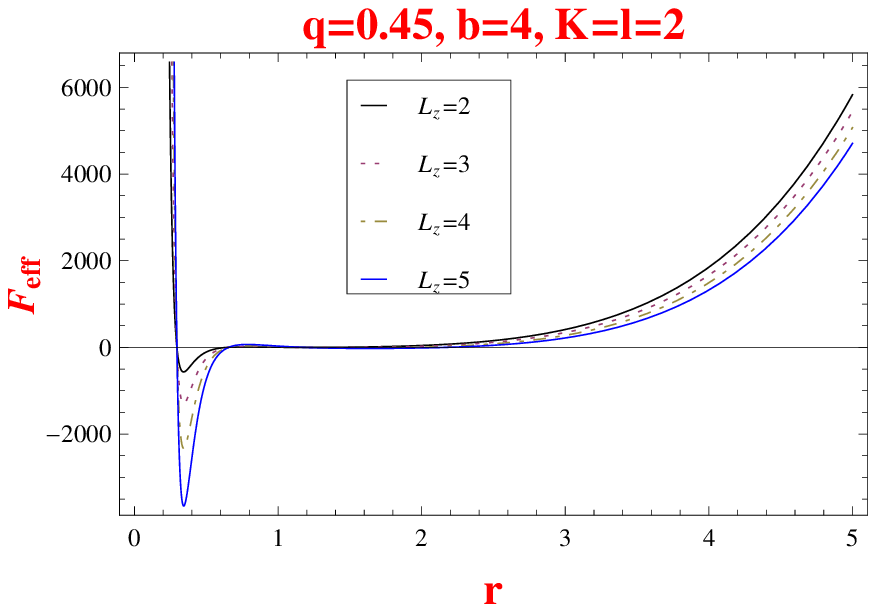}\\
Fig-$ 3c $ represents the variation of effective force $ f_{eff} $  with respect to $ r $ keeping $ q $, $ l $, $ L_z $ and   $ K $ are fixed with varying $b$. \\
Fig-$ 3d $ represents the variation of effective force $ f_{eff} $  with respect to $ r $ keeping $ q $, $ l $, $ K $  and $b$  are fixed with varying $ L_z $.\\
\end{center}
\end{figure}
We have plotted here the graphs of effective force $ f_{eff} $ acting on the charged particle vs the radius $r$ of BH for different values of $q, l, K, b, $ and $L_z$ in Fig- $(3a - 3d)$. We observe from all the graphs that for $r=0.3$ $ f_{eff} $ has very high value, further very small increases of $r$ $, f_{eff} $ decreases sharply and reaches a minimum value at $r=0.4$. $ f_{eff} $ acquires zero value i.e., constant effective potential at about $r=0.6$. Fig-$3a$ and F-g-$3c$ are in similar nature. Both the graphs show for higher values of $q$ and $l$ (Fig-$3a$) and $b$ (Fig-$3c$), $ f_{eff} $ increases more rapidly with increases of $r$. Again surprisingly the feature of Fig-$3b$ and Fig-$3d$ are similar. Here we vary $K$ and $L_z$ respectively. Both these curves show $ f_{eff} $ reaches to Positive value within the range of $r$ about 0.6 to 1.4. For higher values of $K$ and $L_z$ the peak value of $ f_{eff} $ is also high. Beyond $r=1.4$, $ f_{eff} $ decreases slightly and then increases rapidly with increases of $r$ this increasing is higher for higher values of $K$ and $L_z$ respectively. 

\section{Conclusion:}
In our study, we investigate the dynamics of charged particles around charged accelerating AdS (anti-de-Sitter) black hole in the absence and presence of magnetic field by computing $Euler-Lagrange$ equation of motion for radial component only. First of all we choose the charged accelerating AdS (anti-de Sitter) BH with non-linear electromagnetic source and analysis the effective potential, escape energy and therefore escape velocity in the absence and presence of magnetic field. We analysis the dependence of that physical quantities on the radius of BH graphically. We observe from Fig-$1a-1e$ that the escape velocity of the charged particle increases with increase of the radius of BH as expected. But higher values of charge, Ads radius (Fig-$1a$), magnetic field (Fig-$1d$) and escape energy (Fig-$1e$) the escape velocity for same radius decreases. Whereas the escape velocity for same radius decreases initially for higher values of azimuthal angular momentum (Fig-$1b$) and cosmic spring constant (Fig-$1c$). But after a certain range of radius the escape velocity increases for for higher values of azimuthal angular momentum (Fig-$1b$) and cosmic spring constant (Fig-$1c$).  We observe from Fig-$1d$ and Fig-$1e$ that for higher strength of magnetic field as well as escape energy the escape velocity of the particle is also higher. Hence we conclude that the magnetic field which escape the particle from the vicinity of the BH plays an important role in transfer mechanism of energy.

The variation of effective potential with radius of BH is studied in Fig-$2a-2d$. All these graph show that there is a minimum negative value when radius is very small. May be this is a stable circular orbit. There are also may exist many stable circular orbits. Moreover we investigate the CME of two colliding particles with and without magnetic field.

Finally, we calculate the effective force acting on the charged particles in the curved background and examine its dependence on the radius of black bole graphically. We find there is a minimum negative value of effective force at very small radius of BH and then it reaches to zero, stay within few range of the radius for different values of charge and AdS radius (Fig-$3a$) and magnetic field Fig-$3c$). Again for different values of cosmic spring constant (Fig-$3b$) and  azimuthal angular momentum (Fig-$3d$) it reaches to a positive value whose peak value is higher for higher values cosmic spring constant and azimuthal angular momentum separately.

{\bf We have studied the dynamics of both charged and chargeless particles. It is seen that more the charge, lower is the escape velocity. Dependency of escape velocity on $K$ is somewhere different. For the quadratic nature of $K$, we see between two $K_{Low}$ and $K_{high}$ curves, there is a point $r_{crit}$ where the system needs same escape velocity. In a nutshell, it is shown in this paper that as we shift from the Schwarzschild black hole and incorporate charge and magnetic field, the attracting nature increases and particles near to the central engine becomes more redshifted. If we compare our result with existing studies like \cite{Jawad 2016} we observe their speculation is that the particle moving around such black holes should not feel the absence of the singularity of the central regular black hole. The effective part, according to them, must be the magnetic field present in the accretion disc. However, our study also indicates the same. Increment in magnetic field strength signifies that the rotating particle needs comparatively lesser value of escape velocity. Dynamics of a neutral and a charged particle around a Schwarzschild singularity embedded in quintessence universe has been studied in \cite{Jamil 2015}. Energy conditions for stable orbits has been constructed. Their result has been more clearly stated in our work. In future, if the motion of a series of particles are been counted, then we will have an idea of accretion disc around such a regular black hole.} 

\vspace{.1 in}
{\bf Acknowledgment:}
This research is supported by the project grant of Goverment of West Bengal, Department of Higher Education, Science and Technology and Biotechnology (File no:- $ST/P/S\&T/16G-19/2017$). AH wishes to thank the Department of Mathematics, the University of Burdwan for the research facilities provided during the work. RB thanks IUCAA, Pune, India for providing Visiting Associateship.\\\\

\newpage
\vspace{.3in}
\begin{center}
{\Large\bf Appendix}
\end{center}
$\it{Eq^n}$ (\ref{ah2.equn24}) gives 
\begin{equation}\label{ah2.equn47}
A^\mu=\frac{\cal B}{2}\xi_{(\phi)}^\mu~,
\end{equation}
which is nothing but the Maxwell equation for 4-potential $A^\mu$ and where $\xi_{(\phi)}^\mu=\partial_\phi~=(0, 0, 0, 1)$. So among the components $A^t, A^r, A^{\theta}$ and $A^{\phi}$, only $A^{\phi}$ exists and the covariant tensor $A_{\phi}$ is related with the contravariant form as:
\begin{equation}\label{ah2.equn48}
A_{\mu}=g_{\mu\nu}A^\nu~~.
\end{equation}
Now
\begin{equation}\label{ah2.equn49}
A^\phi=\frac{\cal B}{2}\xi_{(\phi)}^\phi=\frac{\cal B}{2}\times1=\frac{\cal B}{2}
\end{equation} 
and 
\begin{equation}\label{ah2.equn50}
\phi=g_{\phi\phi}A^\phi
\end{equation} where
$g_{\phi\phi}$ is given from BH metric (\ref{ah2.equn1}) as,
\begin{equation}\label{ah2.equn51}
g_{\phi\phi}=\frac{g(\theta)r^2sin^2\theta}{\Omega^2}\frac{d{\phi}^2}{K^2}
\end{equation}
Applying $\it{eq^n}$ (\ref{ah2.equn49}) and $\it{eq^n}$ (\ref{ah2.equn51}) in $\it{eq^n}$ (\ref{ah2.equn50}) we obtain,
\begin{equation}
A_\phi=\frac{g(\theta)r^2sin^2\theta}{\Omega^2}\frac{d{\phi}^2}{K^2}\frac{\cal B}{2}
\end{equation}
Accordingly the 2nd part of $\it{eq^n}$ (\ref{ah2.equn28}) turns to be 
\begin{equation}
\frac{q}{m}A_\mu \frac{dx^\mu}{d\tau}=\frac{q}{m} A_\phi \frac{dx^\phi}{d\tau}=\frac{q\cal B}{2m}\frac{g(\theta)r^2sin^2\theta}{\Omega^2}\frac{\dot{\phi}}{K^2}.
\end{equation}
Here $x^\phi$ is actually $\phi$.
\vspace{.2in}


\vspace{.3in}
\begin{thebibliography}{100}

\bibitem{Znajek 1976} Znajek, R. :- {\it Nature} 262, 270 (1976).\
\bibitem{Blandford 1977} Blandford, R. D. and Znajek, R. L. :- {\it Mon. Not. R. Astron. Soc.} {\bf 179}, 433 (1977).\
\bibitem{Frolov 2003} Frolov, V., The Galactic Black Hole, eds. by Falcke, H., Hehl, F. W. IoP, London, (2003).\
\bibitem{Borm 2013} Borm, C. V. and Spaans, M. :- {\it Astron. Astrophys. L} {\bf 9} 553 (2013).\
\bibitem{Mckinney 2007} Mckinney, J. C. and Narayan, R. :- {\it Mon. Not. R. Astron. Soc.} 375, 523 (2007).\
\bibitem{Dobbie 2008} Dobbie, P. B., Kuncic, Z., Bicknell, G. V. and Salmeron, R. :- {\it Proceeding of IAU Symposium 259 Galaxies} Tenerife, Spain (2008).\
\bibitem{Bandos 2009} Ba$n$dos, M., Silk, J. and West, S. M. :-{\it Phys. Rev. Lett.} {\bf 103}, 111102 (2009).\
\bibitem{Berti 2009} Berti, E., Cardoso, V., Gualtieri, L., Pretorius, F. and Sperhake, U. :-{\it Phys. Rev. Lett.} {\bf 103}, 239001 (2009).\
\bibitem{Patil 2010} Patil, M. and Joshi, P. S. :-{\it Phys. Rev. D} {\bf 82}, 104049 (2010).\   
\bibitem{Patil 2011} Patil, M. and Joshi, P. S. :-{\it Classical Quantum Gravity} {\bf 28}, 235012 (2011).\ 
\bibitem{Wei1 2010} Wei, S. W., Liu, Y. X. Guo, H. and Fu, C. E. :-{\it Phys. Rev. D} {\bf 82}, 103005 (2010).\
\bibitem{Zaslavskii 2010} Zaslavskii, O. B. :-{\it JETP Lett.} {\bf 92}, 571 (2010).\
\bibitem{Harada 2011} Harada, T. and Kimura, M. :-{\it Phys. Rev. D} {\bf 83}, 024002 (2011).\
\bibitem{Grib 2011} Grib, A. A. and Pavlov, Yu. V. :-{\it Astropart. Phys.} {\bf 34}, 581 (2011).\
\bibitem{HussianI 2012} Hussian, I. :-{\it Mod. Phys. Lett. A} {\bf 27}, 1250017 (2012).\
\bibitem{Sadeghi 2012} Sadeghi, J. and Pourhassan, B. :-{\it Eur. Phys. J. C} {\bf 72}, 1984 (2012).\
\bibitem{Bambi 2013} Bambi, C. and Modesto, L. :-{\it Phys. Lett. B} {\bf 721}, 329 (2013).\
\bibitem{Tuesunov 2013} Tuesunov, A., Kolo.s, M., Abdujabbarov, A. Ahmedov, B. and Syuchlik, Z. :-{\it Phys. Rev. D} {\bf 88}, 124001 (2013).\
\bibitem{Harada 2014} Harada, T. and Kimura, M. :-{\it Classical Quantum Gravity} {\bf 31}, 243001 (2014).\
\bibitem{Lake 2010} Lake, K. :-{\it Phys. Rev. Lett.} {\bf 104}, 211102 (2010).\ 
\bibitem{Wei2 2010} Wei, S. W., Liu, Y.X., Li, H.T. and Chen, F. W. :-{\it JHEP} {\bf 12}, 066 (2010).\
\bibitem{Liu 2011} Liu, C., Chen, S., Ding, C. and Jing, J. :-{\it Phys. Lett. B} {\bf 701}, 285 (2011).\
\bibitem{Said 2011} Said, J. L. and Adami, K.Z. :-{\it Phys. Rev. D} {\bf 83}, 104047 (2011).\
\bibitem{Sharif 2013} Sharif, M.and Haider, N.  :-{\it Theor, J. Exp. Phys.} {\bf 117}, 78 (2013).\
\bibitem{Zakria 2015} Zakria, A. and Jamil, M. :-{\it JHEP} {\bf 147}, 1505 (2015).\
\bibitem{Pradhan 2015} Pradhan, P. :-{\it Astropart. Phys.} {\bf 62}, 217 (2015).\
\bibitem{Forlov 2010} Forlov, V. P. and Shoom, A. A. :-{\it Phys. Rev. D} {\bf 82}, 084034 (2010).\
\bibitem{Forlov 2012} Forlov, V. P. :-{\it Phys. Rev. D} {\bf 85}, 024020 (2012).\
\bibitem{Zahrani 2013} Zahrani, A. A., Frolov, V. P. and Shoom, A. A. :-{\it Phys. Rev. D} {\bf 87}, 084043 (2013).\
\bibitem{Nakamura 1993} Nakamura, Y., and Ishizuka, T. :-{\it Astrophys. Space Sci.} {\bf 210}, 105 (1993).\
\bibitem{Takahashi 2009} Takahashi, M. and Koyama, H. :-{\it Astrophys. J} {\bf 693}, 472 (2009).\
\bibitem{Kopacek 2010} Kopacek, O., Kovar, J., Karas, V. and Stuchlik, Z. :-{\it AIP Conf. Proc.} 1{\bf 283}, 278 (2010).\
\bibitem{Preti 2010} Preti, G. :-{\it Phys. Rev. D} {\bf 81}, 024008 (2010).\
\bibitem{Forlov 2011} Forlov, V. P. and Krtous, P. :-{\it Phys. Rev. D} {\bf 83}, 024016 (2011).\
\bibitem{Igata 2011} Igata, T., Koike, T. and Ishihara, H. :-{\it Phys. Rev. D} {\bf 83}, 065027 (2011).\
\bibitem{Pugliese 2011} Pugliese, D., Quevedo, H. and Rufini, R. :-{\it Phys. Rev. D} {\bf 83}, 104052 (2011).\
\bibitem{Majeed 2015} Majeed, B., Hussain, S. and Jamil, M. :-{\it Advances in High Enegy Physics 2015}, 6712559 (2015).\
\bibitem{Jamil 2015} Jamil, M., Hussian, S. and Majeed, B. :-{\it Eur. Phys. J. C} {\bf 75}, 24 (2015).\
\bibitem{Jawad 2016} Jawad, A., Ali, F., Jamil, M. and Debnath, U. :-{\it \it  arXiv} {\bf 1610.07411}, [gr-qc] (2016).\
\bibitem{Zhou_added} Zhou, S. et al. :- {\it  International Journal of Theoretical Physics (IJTP)} {\bf 54}, 2905(2015)\
\bibitem{Chen_added} Chen, J. H. et al :- {\it International Journal of Modern Physics A. (IJMPA)} {\bf 25 No. 7} 1439(2010).\
\bibitem{Griffith 2006} Griffith, J. B. and  Podolsky, J. :-{\it A new look at the plebanski-demianski family of solutions, Int. J. Mod. Phys. D} {\bf 15}, 335-370 (2006).\
\bibitem{Hang 2016} Hang Liu and Xin-he Meng :- {\it  arXiv}: {\bf 1607.00496}, [gr-qc] (2016).\
\bibitem{Wald 1974} Wald, R. M :- {\it Phys Rev D} {\bf 10}, 1680(1974).\
\bibitem{Aliev 1978} Aliev, A. N. and Ozdemir, N. :- {\it Monthly Notices of the Royal Astronomical Socity (MNRAS)} {\bf 336}, 241(1978).\
\bibitem{Hsu 2004} Hsu, S. D. H. :-{\it Phys. Lett. B} {\bf 594}, 13 (2004).\

\bibitem{HussainS 2014} Hussain, S., Hussain, I. and Jamil, M. :-{\it Eur. Phys. J. C} {\bf 74}, 3210 (2014).\
\end{thebibliography}
\end{document}